\documentclass[11pt]{article}
\usepackage{amsfonts,amssymb,amsfonts,amsgen,amsmath}

\def\i{\begin{eqnarray}}\def\f{\end{eqnarray}}
\def\del#1#2{\frac{\partial{#1}}{\partial {#2}}}
\def\d{\partial}\def\C{{\mathbb C}}
\def\q{\quad}\def\R{{\mathbb R}}\def\Z{{\mathbb Z}}\def\L{\Lambda}\def\vd{\delta}
\def\ep{\epsilon}\def\l{\lambda}\def\th{\theta}\def\lan{\langle}\def\ran{\rangle}
\def\hW{\hat{W}}\def\hw{\hat{w}}\def\tt{\hat{t}}
\def\ss{\hat{s}}\def\ot{\otimes}
\def\hxi{\hat{\xi}}\def\heta{\hat{\eta}}
\def\A{\mathcal{A}}\def\F{\mathcal{F}}

\begin{document}

\begin{center}
{\Large\bf
Factorization methods for noncommutative KP \\ \bigskip
and Toda hierarchy}\\ \bigskip\bigskip
Masato Sakakibara \\ \smallskip
\textit{\small Department of Physics, University of Tokyo, \\ \smallskip
Tokyo 113-0033, Japan \\ \smallskip
sakakiba@monet.phys.s.u-tokyo.ac.jp}
\end{center}
                                                                                
\begin{abstract}
We show that the solution space of the noncommutative KP hierarchy 
is the same as that of the commutative KP hierarchy owing to the
Birkhoff decomposition of groups over the noncommutative algebra.
The noncommutative Toda hierarchy is introduced. 
We derive the bilinear identities for the Baker--Akhiezer functions
and calculate the $N$-soliton solutions of the noncommutative Toda
hierarchy.
\end{abstract}

\section{Noncommutative KP and Toda hierarchy}

Recently, the classical and quantum field theories over 
the noncommutative (NC) space-times have been extensively studied. 
The NC gauge theory has had a great success in the string theory. 
In particular, the NC deformation of the ADHM construction 
of the (anti)SDYM equation \cite{nek} and the Nahm construction \cite{bak} were shown.
It seems to imply the significance of the NC deformations of the integrable systems. 
Some authors have discussed the NC integrable systems on 
lower dimension \cite{dim,ham,lec,shi,wan} (see also \cite{kup,tak2,tak3}), 
especially the NC KP hierarchy. In this letter, we study the moduli space of the 
NC KP hierarchy throughout the Birkhoff decomposition
of a certain formal group over a NC algebra.
We also introduce the NC Toda hierarchy and derive
the bilinear identities and the $N$-soliton solutions.

Let $(\tt_1,\tt_2,\cdots)$ be coordinates of NC plane $\R^{2\infty}$ 
which satisfy $[\tt_n,\tt_m]=i\th_{nm}$, and $\A_\th$ a set of functions on it. 
The deformation parameters $\th_{nm}\in\R$ are non-zero 
constants and we assume that a matrix $(\th_{nm})$ is invertible. 
By an orthogonal change of coordinates as
$[\tt_{2n-1},\tt_{2n}]=i\th_n\; (n\ge1)$, the algebra is realized as operators
over the Fock space 
$\F=\oplus_{n_1,n_2,\cdots}{\mathbb C}|n_1n_2\cdots\ran$ $(n_i=0,1,\cdots)$. 

The NC KP hierarchy of the operator form is defined as follows \cite{dim,ham,wan}.
We consider an operator valued monic pseudo differential
operator\footnote{The variable $x$ needs not to be an operator on $\F$,
since the Lax equations give $\del{\hat{u}_n}{x}=\del{\hat{u}_n}{\tt_n}$
for the Lax operator $L=\sum_{n\ge-1}\hat{u}_{-n}\d^{-n}$.} (PDO)  
\i \hW={\bf 1}_\F+\sum_{n\ge1}\hw_n(x,\hat{t})\d^{-n}_x,  \f
where ${\bf 1}_\F$ is an identity of $\A_\th$.
The coefficients of $\hW$ are operators  
\i \hw_n(x,\hat{t})=\sum_{l_i,m_i\ge0}w_n^{(l,m)}(x)
|l_1l_2\cdots\ran\lan m_1m_2\cdots|, \f
where $w_n^{(l,m)}$ are ${\mathbb C}$-valued functions of $x$.
The Sato--Wilson equations of the NC KP hierarchy are given by 
\i \label{sw}
\del{}{\hat{t}_n}\hW=-(\hW\d^n_x\hW^{-1})_-\hW\q(n\ge1), \f
where\footnote{This notation is different from
$\del{}{\tt_n}=\hat{\d}_n=-i\sum_{m}\th_{nm}^{-1}\tt_m$
which is often used in NC gauge theory.} 
$\del{}{\hat{t}_n}:=-i\sum_m\th_{nm}^{-1}[\hat{t}_{m},\cdot]$.
For a PDO $P$, $(P)_+$ denotes a part of a differential operator of $P$
and $(P)_-$ denotes $P-(P)_+$.
\vspace{3mm}

Here we introduce the NC Toda hierarchy. 
We add coordinates $(\ss_1,\ss_2,\cdots)$  
which satisfy $[\ss_n,\ss_m]=i\tilde{\th}_{nm},\; [\ss_n,\tt_m]=0$ 
and $\A_{\th,\tilde{\th}}$ denotes a set of functions on it.
This algebra is realized over a tensor product of
two Fock spaces $\F\otimes\F$. 
Let $\hW^{(+)}$ (resp. $\hW^{(-)}$) be an upper (resp. lower)
triangular $\Z\times\Z$ matrix
whose components are valued in $\A_{\th,\tilde{\th}}$. 
All diagonal components of the matrix $\hW^{(-)}$ are an identity  
${\bf 1}_{\F\ot\F}$ of $\A_{\th,\tilde{\th}}$. 
We define the Sato--Wilson equations of the NC Toda hierarchy as 
\i \label{toda} 
&&\del{\hW^{(\pm)}}{\tt_n}=\pm(\hW^{(-)}\L^{n}\hW^{(-)-1})_\pm\hW^{(\pm)},
\nonumber\\
&&\del{\hW^{(\pm)}}{\ss_n}=\mp(\hW^{(+)}\L^{-n}\hW^{(+)-1})_\pm\hW^{(\pm)}, 
\f 
for $n\ge1$, where $(n,m)$ component of $\L$ is 
$\vd_{n+1,m}{\bf 1}_{\F\ot\F}$ and
$\del{}{\ss_n}:=-i\sum_m\tilde{\th}_{nm}^{-1}[\ss_{m},\cdot]$.
For a matrix $P$, $(P)_+$ denotes the upper triangular part of $P$
and $(P)_-$ denotes $P-(P)_+$.
Note that these equations give the Lax equations 
and the Zakharov--Shabat (ZS) equations as usual \cite{tak1}. We define 
$\hat{b}_n,\hat{c}_n\in \A_{\th,\tilde{\th}}$ as
$B:=(\hW^{(-)}\L\hW^{(-)-1})_+=\L+\sum_{n\in\Z}\hat{b}_nE_{nn}$ and 
$C:=(\hW^{(+)}\L^{-1}\hW^{(+)-1})_-=\sum_{n\in\Z}\hat{c}_nE_{nn-1}$
where $(k,l)$ component of $E_{nm}$ is $\vd_{kn}\vd_{lm}$. 
The ZS equation $\del{B}{\ss_1}-\del{C}{\tt_1}+[B,C]=0$
gives the NC Toda equations
\i \del{\hat{b}_n}{\ss_1}=-\hat{c}_{n+1}+\hat{c}_n,\q 
\del{\hat{c}_n}{\tt_1}=\hat{b}_n\hat{c}_n
-\hat{c}_n\hat{b}_{n-1}, \f
for $n\in\Z$. In the commutative case we can introduce functions $\phi_n$ such that 
$\hat{b}_n=-\del{}{t_1}\phi_n$ and $\hat{c}_n=e^{\phi_{n-1}-\phi_n}$. 
However, in the NC case we cannot define such functions.

\section{A solution space of the NC KP hierarchy}

First we consider the NC KP hierarchy.
As in the commutative case, the Sato--Wilson equations (\ref{sw})
are equivalent to the Birkhoff decomposition of a certain formal group $G$.
The integrable hierarchy throughout the factorization problems 
$G=G_-G_+$ of a group $G$ over a NC algebra $R$
was studied in \cite{mul1,tak2}, for example (see also \cite{mul2}).
In our case $G,G_\pm$ correspond to the formal groups
$\exp{\mathcal G},\exp{\mathcal G}_\pm$ in \cite{tak2} 
over the NC plane $\R^{2\infty}$ instead of $\R^2$ (see remark 1.). 
 
Let $\hW_0\in G_-$ be an `initial value' of $\hW$ which is a monic PDO
and satisfies $\del{}{\tt_n}\hW_0=0$ for $n\ge1$.
The factorization of $e^{\hxi}\hW_0^{-1}$ 
\i \label{fact}
e^{\hxi}\hW_0^{-1}=\hW^{-1}\cdot \hW^{(+)}\q\q (\hW\in G_-,\; \hW^{(+)}\in G_+), \f
where $\hxi=\sum_{n\ge1}\tt_n\d^n_x$ gives 
the Sato--Wilson equations (\ref{sw}) as in the commutative case \cite{mul2,mul1,tak1}. 
Conversely, we assume that $\hW$ satisfies the Sato--Wilson equations,
the factorization 
\i \label{con}
e^{-\hxi}\hW^{-1}=\hat{M}^{(-)-1}\hat{M}^{(+)}\q\q
(\hat{M}^{(\pm)}\in G_\pm), \f 
gives equations 
\i -\hat{M}^{(+)}(\hW\d^n_x\hW^{-1})_+\hat{M}^{(+)-1}=
-\del{\hat{M}^{(-)}}{\tt_n}\hat{M}^{(-)-1}+\del{\hat{M}^{(+)}}{\tt_n}
\hat{M}^{(+)-1}, \f
for $n\ge1$. Taking a part of $(\cdot)_-$
of both sides, we obtain $\del{}{\tt_n}\hat{M}^{(-)}=0$. Since 
$e^{\hxi}\hat{M}^{(-)-1}=\hW_-^{-1}\hat{M}^{(+)-1}$ 
gives the factorization of $e^{\hxi}\hat{M}^{(-)-1}$ 
by the uniqueness of the Birkhoff decomposition,
we obtain equation (\ref{fact}).

In the noncommutative case, $\{\tt_n\}$ are operators and 
the fact $\del{}{\tt_n}\hW_0=0$ has a quite different meaning from that of
the commutative case.
Since the derivative of $\tt_n$ is defined as $\del{}{\tt_n}=-i\sum_m\th_{nm}^{-1}[\tt_m,\cdot]$, 
it means that $\hW_0$ commutes with all $\tt_n$ for
$n\ge1$. Such an operator must be proportional to the identity of
$\A_\th$, and $\hW_0$ must have a 
form\footnote{The coefficient functions $w^{(0)}_n$ possibly depend on
$\th_{nm}$. However they only shift the initial value and are not
intrinsic to the NC KP hierarchy.}
\i \label{ini}
\hW_0={\bf 1}_\F\left(1+\sum_{n\ge1}w_n^{(0)}(x)\d^{-n}_x\right). \f
Therefore $\hW_0$ defines a point of the solution space of 
the commutative KP hierarchy (so-called the Sato Grassmanian \cite{sat} and so on).
Since the operator $\hW$ corresponds to $\hW_0$ uniquely through the action of
operator $e^{\hxi}$, we conclude that, in this sense,
the solution space of NC KP hierarchy is the same as the commutative one.
This fact is valid for the multicomponent case and their reductions.
In general, for a NC hierarchy obtained by the Birkhoff decomposition, 
such as the NC Toda hierarchy, the solution space is same with 
the space of the initial values.
\vspace{5mm}

\noindent {\bf Remark 1}\q
In \cite{mul1}, they studied the group $G=\hat{\mathcal E}^\times$ which is a set of 
PDOs $P$ whose coefficients are valued in $R[[t]]$ with time parameters $t=\{t_1,t_2,\cdots\}$ 
under the conditions $P|_{t=0}$. In the case of the NC KP hierarchy, $\{\tt_n\}$ are elements of
$R=\A_\th$ and we cannot adopt their results to our case directly.
Considering the Neumann series (see the proof of Theorem 3.2 \cite{mul1}), we obtain the Birkhoff
decomposition of $G$ as formal series $\C[[\tt]]=\C[[\tt_1,\tt_2,\cdots]]$. 
The problem is whether components are well-defined under the re-arrangement
of generators $\{\tt_n\}$.
However, substituting ${\rm val}_t(\th_{nm}):=n+m$ by the relations $[\tt_n,\tt_m]=i\th_{nm}$,
we find that components are well-defined as a
formal power series of $\A_\th[[\th_{nm}]]$  from theorem 3.2 \cite{mul1}.
Note that we need the assumption that the LHS of (\ref{con}) 
is an element of $\hat{\mathcal E}^\times$ \cite{mul1} with replacing 
operators $\tt_n$ to time variables $t_n$ for its factorization.
\vspace{5mm}

\noindent
{\bf Remark 2}\q In the limit $\th_{nm}\longrightarrow 0$, 
the NC KP hierarchy is reduced to
the commutative one and $\hw_n$ to
\i w_n(t,x)=\lim_{\th\longrightarrow 0} \sum_{l_i,m_i\ge 0}w_n^{(l,m)}
(x)f_{(l,m)}(t,x,\th), \f
with commutative times $t=(t_1,t_2,\cdots)$,
where $f_{(l,m)}$ are functions obtained by
the Wigner--Weyl transformation (see for e.g., \cite{har}) of 
$|l_1l_2\cdots\ran\lan m_1m_2\cdots|$.
The point of the solution space where $w_n$ are defined
is same as that of $\hw_n$.
\vspace{5mm}

\noindent
{\bf Remark 3}\q In \cite{dim}, they derive the 
differential equations of the deformation parameters $\th_{nm}$,
and remark that those equations allow us to 
construct the solutions of the NC KP hierarchy as a formal
Taylor series of $\th_{nm}$ from the solutions of the commutative KP hierarchy.

\section{Noncommutative Toda hierarchy}

In this section, we define the Baker--Akhiezer (BA) functions for the NC
Toda hierarchy, and show that they satisfy the bilinear identities. We
also calculate the $N$-soliton solutions.

\subsection{Bilinear identities}

Let $\hw_{\pm n}(s),\hw_{\pm n}^\ast(s)\in \A_{\th,\tilde{\th}}$
be operators 
which are coefficients of $\hW^{(\pm)}$ and $\hW^{(\pm)-1}$
\i \label{coe}
\hW^{(\pm)}=\sum_{n\ge0}{\rm diag}({\hw_{\pm n}(s)})\L^{\pm n}, \q
\hW^{(\pm)-1}=\sum_{n\ge0}\L^{\pm n}{\rm diag}({\hw^\ast_{\pm n}(s)}) \f
where ${\rm diag}(\hw_{+n}(s))
={\rm diag}(\cdots,\hw_{+n}(-1),\hw_{+n}(0),\cdots)$ and so on.
We introduce the operator valued Baker--Akhiezer functions $\hw_\pm,\;\hw_\pm^\ast$
\i 
&&\hw_-(s,\l)=\left(\sum_{n\ge0}\hw_{-n}(s)\l^{-n+s}\right)e^{\hxi_-},\q   
\hw_+(s,\l)=\left(\sum_{n\ge0}\hw_{+n}\l^{n+s}\right)e^{\hxi_+}, \nonumber\\
&&\hw_-^\ast(s,\l)=e^{-\hxi_-}\left(\sum_{n\ge0}\hw_{-n}^\ast(s)\l^{-n-s}\right),\q   
\hw_+^\ast(s,\l)=e^{-\hxi_+}\left(\sum_{n\ge0}\hw_{+n}^\ast\l^{n-s}\right), 
\f
where $\hxi_-:=e^{\sum_{n\ge1} \tt_n\l^n}$ and $\hxi_+:=e^{\sum_{n\ge1} \ss_n\l^{-n}}$.
Considering the same arguments with the commutative case \cite{tak1}, 
we obtain the bilinear identities as follows.
By the definition of $\hw_\pm,\hw^\ast_\pm$ (\ref{coe}), the BA functions satisfy
\i \label{bi1}
\oint\frac{d\l}{2\pi i} 
\hw_-(s',\l)\hw_-^\ast(s,\l)=\oint \frac{d\l}{2\pi i} 
\hw_+(s',\l)\hw^\ast_+(s,\l)(={\bf 1}_{\F\ot\F}\vd_{s',s-1}).  \f
Since $\hw_+$ and $\hw_-$ satisfy the same linear differential equations of $\tt_n$
and $\ss_n$,  
\i\del{\hw_\pm}{\tt_n}=(\hW^{(-)}\L^n\hW^{(-)-1})_+\hw_\pm,\q  
\del{\hw_\pm}{\ss_n}=(\hW^{(+)}\L^{-n}\hW^{(+)-1})_-\hw_\pm, \f
there exist an operator $\hat{B}_{IJ}$ such that
$\d_{\tt}^I\d_{\ss}^J\hw_\pm= B_{IJ}\hw_\pm$
where $\d_{\tt}^I=\del{}{\tt_{i_1}}\cdots\del{}{\tt_{i_l}}$ and so on.
We multiply $\hat{B}_{IJ}$ on both sides of (\ref{bi1})
with the identification $\L$ with $e^{\del{}{s'}}$ for $\hw_\pm(s',\l)$ \cite{tak1},
and obtain the bilinear identities 
\i \label{bi2}
\oint\frac{d\l}{2\pi i} 
\d_{\tt}^I\d_{\ss}^J\hw_-(s',\l)\cdot\hw_-^\ast(s,\l)
=\oint\frac{d\l}{2\pi i} 
\d_{\tt}^I\d_{\ss}^J\hw_+(s',\l)\cdot\hw^\ast_+(s,\l),  \f
for any $I,J$. 
For $\d_{\tt}^I\d_{\ss}^J=\del{}{\tt_n}$ (resp. $\d_{\tt}^I\d_{\ss}^J=\del{}{\ss_n}$), 
equations (\ref{bi2}) are equal to
$(\hW^{(-)}e^{n\del{}{s'}}\hW^{(-)-1})_+\vd_{s',s-1}$ (resp. $(\hW^{(+)}e^{-n\del{}{s'}}\hW^{(+)-1})_-\vd_{s',s-1}$) 
and the bilinear identities are equivalent to the Sato--Wilson equations (\ref{toda}) as
in the commutative case. Note that we cannot introduce the notion of the
$\tau$-functions \cite{sat,tak1} by the same reason with the NC KP
hierarchy \cite{dim}.
If we set $s=s'$, the RHS of (\ref{bi2}) is zero and these identities correspond to those
of the NC KP hierarchy \cite{dim}. This fact means that if we 
restrict the Fock space $\F\ot\F$ to $\F$ on which $\tt_n$ act, 
the BA functions of the NC Toda hierarchy for fixed $s$ are also those of
the NC KP hierarchy.

\subsection{$N$-soliton solutions}
In this section we calculate $N$-soliton solutions for the NC Toda
hierarchy considering the factorization problem. We put
$\hat{V}^{(-)}:=\hW^{(-)}e^{-\sum_{n\ge1}\ss_n\L^{-n}}$ and 
$\hat{V}^{(+)}:=\hW^{(+)}e^{-\sum_{n\ge1}\tt_n\L^n}$. 
Then, the Sato--Wilson equations (\ref{toda})
are equivalent to the factorization \cite{tak1}
\i \label{gauss}
\hat{V}^{(-)-1}\cdot\hat{V}^{(+)}=e^{\hxi}\hat{V}_0^{-1}e^{-\hxi}.
\f
where $\hxi:=\exp(\sum_{n\ge1}\tt_n\L^n+\ss_n\L^{-n})$.
We choose the initial value for the $N$-soliton solutions (cf (\ref{ini})) 
\i \hat{V}_0^{-1}={\bf 1}_{\F\ot\F}\left(
I+\ep\sum_{i=1}^Na_i\sum_{m,n\in\Z}p^m_iq^{-n}_i E_{mn}\right), \f
where $a_i,p_i,q_i$ are positive constant parameters 
such that $q_N<\cdots<q_1<p_1<\cdots<p_N$,
$\ep$ is a formal parameter, and $I:=\sum_{n\in\Z} E_{nn}$.
We put $\hat{V}^{(-)}={\bf 1}_{\F\ot\F}I+\hat{Z},\; \hat{Z}=(\hat{z}_{ij})$ where 
$\hat{z}_{nm}\in A_{\th,\tilde{\th}}$ is zero for $n\le m$.
Considering equation (\ref{gauss}) we have
\i 
&&(\cdots,\hat{z}_{s,s-2},\hat{z}_{s,s-1})\left(1_{\F\ot\F}I+\ep\sum_{i=1}^N a_i 
e^{\heta(p_i)}{\boldsymbol p}_i{\boldsymbol q}_ie^{-\heta(q_i)}
\right)\nonumber\\
&&\hspace{6cm}=-\ep\sum_{j=1}^Na_j p_j^s 
e^{\heta(p_j)}{\boldsymbol q}_je^{-\heta(q_j)}, \f
where ${\boldsymbol p}_i:={}^t(\cdots,p_i^{s-2},p_i^{s-1}),\;
{\boldsymbol q}_i:=(\cdots,q_i^{-s+2},q_i^{-s+1})$ and 
$\heta(p):=\sum_{n\ge1}(\tt_np^n+\ss_np^{-n})$.
In the commutative case, the Cramer
formula solves these equations and solutions are expressed by the $\tau$-function.
In the NC case we cannot adopt such a method. 
We solve this equation as a formal 
series
\footnote{Since it reduces to the soliton solutions of the NC KP
hierarchy and depends only on $\tt_n p^n+\ss_n p^{-n}\; (p\in\{p_i,q_i\})$
for $\tt_n,\ss_n$, we find that this series is well-defined as an element of
$\A_{\th,\tilde{\th}}[[\th_{nm},\tilde{\th}_{nm}]]$ without $\ep$.} 
of $\ep$, 
\i &&(\cdots,\hat{z}_{s,s-2},\hat{z}_{s,s-1})=
-\ep\sum_{j=1}^Na_j p_j^s e^{\heta(p_j)}{\boldsymbol q}_je^{-\heta(q_j)}
\sum_{k\ge0}\left(-\ep\sum_{i=1}^N\ a_i 
e^{\heta(p_i)}{\boldsymbol p}_i{\boldsymbol q}_ie^{-\heta(q_i)}\right)^k
\nonumber\\
&& =\sum_{k\ge0}(-\ep)^{k+1}\hspace{-5mm}
\sum_{1\le j,i_1,\cdots,i_k\le N} \hspace{-5mm} p_j^s 
\hat{\phi}_j\hat{\phi}_{i_1}\cdots\hat{\phi}_{i_k}
\frac{q_j^{-s+1}p_{i_1}^{s}}{p_{i_1}-q_{j}}
\frac{q_{i_1}^{-s+1}p_{i_2}^{s}}{p_{i_2}-q_{i_1}}\cdots
\frac{q_{i_k}^{-s+1}p_{i_{k-1}}^{s}}{p_{i_{k}}-q_{i_{k-1}}}{\boldsymbol q}_{i_{k}},
\f
where $\hat{\phi}_i:=a_ie^{\heta(p_i)}e^{-\heta(q_i)}$.
Here we used ${\boldsymbol q}{\boldsymbol p}=\frac{q^{-s+1}p^s}{p-q}$.
Thus we obtain the $N$-soliton solution of the NC Toda hierarchy,
\i &&\hat{z}_{nm}=-\sum_{k\ge 0}\ep^{k+1}\hspace{-5mm}
\sum_{1\le i_0,i_1,\cdots,i_k\le N}\hspace{-5mm}
q_{i_k}^{-m+n-1}\frac{(p_{i_0}\cdots p_{i_k})^n}{(q_{i_0}\cdots
q_{i_k})^{n-1}} \nonumber\\
&&\hspace{5cm}
\times\frac{\hat{\phi}_{i_0}\hat{\phi}_{i_1}\cdots\hat{\phi}_{i_k}}
{(q_{i_0}-p_{i_1})(q_{i_1}-p_{i_2})\cdots(q_{i_{k-1}}-p_{i_k})},
\f
for $n>m$. As noted in previous section, if we restrict the Fock
space $\F\ot\F$ to $\F$ on which $\tt_n$ act,  
the BA functions of the NC Toda hierarchy are reduced to those of the NC KP hierarchy. 
In fact, the solution with a change of parameters $a_i\longrightarrow a_i/p_i$ 
\i \phi:=-\hat{z}_{10}=\sum_{k\ge 0}\ep^{k+1}\hspace{-2mm}
\sum_{1\le i_0,\cdots,i_k\le N}\frac{\hat{\phi}_{i_0}\cdots\hat{\phi}_{i_k}}
{(q_{i_0}-p_{i_1})\cdots(q_{i_{k-1}}-p_{i_k})},  \f
corresponds the $N$-soliton solutions of the NC KP hierarchy 
obtained in \cite{dim,pan} using the trace method \cite{ohk}.


\begin{thebibliography}{99}

\bibitem{bak} D. Bak, Phys. Lett. {\bf B471}, 149 (1999).
\bibitem{dim} A. Dimakis, F. Muller-Hoissen, Jour. Phys. {\bf A37} (2004)
	10899 [hep-th/0406112].
\bibitem{ham} M. Hamanaka, K. Toda, Jour. Phys. {\bf A36}, 11981 (2003); 
	      M. Hamanaka, hep-th/0311206.
\bibitem{har} J. A. Harvey, hep-th/0102076.
\bibitem{kup} B. A. Kupershmid, {\it KP or mKP: noncommutative mathematics
	of Lagrangian, Hamiltonian and integrable systems}, American Mathematical Society, 2000.
\bibitem{lec} O. Lechtenfeld, L. Mazzanti, S.Penati, A.D. Popov,
	L. Tamassia, Nucl. Phys. {\bf B705} (2005) 477 [hep-th/0406065].
\bibitem{mul2} M. Mulase, Inv. Math, {\bf 54}, 57 (1984).
\bibitem{mul1} M. Mulase, Inv. Math, {\bf 92}, 1 (1988).
\bibitem{nek} N. Nekrasov,  A.Schwarz, Comm. Math. Phys. {\bf 198}, 689 (1998).
\bibitem{ohk} K. Ohkuma , M. Wadati, Jour. Phys. Soci. Jpn. {\bf 52}, 749
	(1983).
\bibitem{pan} L. D. Paniak, hep-th/0105185. 
\bibitem{sat} M. Sato, RIMS Kokyuroku {\bf 439}, 30 (1981); M.Sato,
	Y. Sato, ``Soliton equations as dynamical systems on infinite
	dimensional Grassmann manifold'', in P.D. Lax, H. Fujita, and G. Strang (eds.), 
	{\it Nonlinear Partial Differential Equations in Applied
	Sciences}, (North-Holland, Amsterdam 1982).
\bibitem{shi} K. Shigechi, N. Wang, M. Wadati, Nucl. Phys. {\bf B706} (2005)
	518 [hep-th/0404249]. 
\bibitem{tak1} K. Takasaki, K. Ueno, Adv. Stud. Pure. Math. {\bf 4}, 1
	(1994);  
\bibitem{tak2} K. Takasaki, J. Geom. Phys. {\bf 14}, 111 (1994); 
               K. Takasaki, J. Geom. Phys. {\bf 14}, 332 (1994).
\bibitem{tak3} K. Takasaki, J. Geom. Phys. {\bf 37}, 291 (2001).
\bibitem{wan} N. Wang, M. Wadati, J. Phys. Soc. Jpn. {\bf 72}, 1366;
              N. Wang, M. Wadati, J. Phys. Soc. Jpn. {\bf 72}, 1881;     
              N. Wang, M. Wadati, J. Phys. Soc. Jpn. {\bf 72}, 3055;
              N. Wang, M. Wadati, J. Phys. Soc. Jpn. {\bf 73}, 1689.    
\end{thebibliography}
\end{document}